\documentclass[aps,prd,showpacs,floats,letterpaper,floatfix,
groupedaddress,superscriptaddress
,nofootinbib
,eqsecnum
,twocolumn
]{revtex4}
\usepackage{amsmath}
\usepackage{amsfonts}
\usepackage{bm}
\usepackage{color}
\usepackage{graphicx}

\def\v1v2{{\bf v}_1 \cdot {\bf v}_2}
\def\bx{\bm{x}}
\allowdisplaybreaks
\graphicspath{ {./images/}}
\begin{document}

\title{Higher-order effects in the dynamics of hierarchical triple systems. III.  Astrophysical implications of second-order and dotriacontapole terms}

\author{
Landen Conway} \email{conwayl@ufl.edu}
\affiliation{Department of Physics, University of Florida, Gainesville, Florida 32611, USA}

\author{
Clifford M.~Will} \email{cmw@phys.ufl.edu}
\affiliation{Department of Physics, University of Florida, Gainesville, Florida 32611, USA}
\affiliation{GReCO, Institut d'Astrophysique de Paris, CNRS,\\ 
Universit\'e Pierre et Marie Curie, 98 bis Boulevard Arago, 75014 Paris, France}

\date{\today}

\begin{abstract}
We study the long-term evolution of selected hierarchical triple systems in Newtonian gravity.  We employ analytic equations derived in Paper II for the evolution of orbit-averaged orbital elements for both inner and outer orbits, which include two classes of contributions.  One class consists of linear-order contributions, including quadrupole, octupole, hexadecapole and dotriacontapole orders, the latter scaling as $\epsilon^6$, where $\epsilon = a/A$, the ratio of the semimajor axes of the inner and outer orbits.  The second class consists of contributions at {\em second} order in the fundamental perturbation parameter; they contribute at orders $\epsilon^{9/2}$, $\epsilon^{5}$, $\epsilon^{11/2}$ and $\epsilon^{6}$. For well studied triples such as star-planet systems perturbed by a low-mass third body (``hot Jupiters''), second-order and dotriacontapole (SOD) effects induce only small corrections.  For stellar-mass binaries orbiting supermassive black holes, SOD corrections can suppress orbital flips that are generated by purely first-order effects.  Planets orbiting binary star systems are susceptible to significant variations in the planetary semimajor axis, an effect that does not occur at first perturbative order.  SOD effects in triple black hole systems can induce migrations of the eccentricity to significantly larger values than predicted by first-order perturbations, with implications for the gravitational-wave induced inspiral of the inner binary.  We also show that in most cases, evolutions using our SOD equations are in better agreement with those from direct integration of the N-body equations of motion than those from first-order perturbations through hexadecapole order.

\end{abstract}

\maketitle

\section{Introduction and summary}
\label{sec:intro}

This is the third in a series of papers that address the dynamics of hierarchical triple systems, stellar systems in which two bodies orbit each other in the presence of a distant third body.  Although research on these systems has a rich history dating back to Newton himself, current work is very active, with applications in exoplanet science, evolution of stars and stellar clusters, solar system and planetary science, and gravitational-wave astronomy (for reviews see \cite{2016ARA&A..54..441N,2013degn.book.....M,2006tbp..book.....V,2008LNP...760...59M,1999ssd..book.....M}).  

Our approach to this problem has been to adopt the Lagrange planetary equations (LPE) to describe the evolution of the ``osculating orbit elements'' of the  pair of effective orbits that describe the three-body system in a basis of coordinates centered on the system's center of mass.  One is the relative orbit between the bodies of the inner binary and the other is the relative orbit between the outer body and the center of mass of the inner binary.  While many textbooks give formal and long-winded derivations of the orbit elements and the LPE, fundamentally they are nothing but an exact change of variables from Newton's equations in terms of $\bm x$ and $\bm v$ of each orbit to an equivalent and exact set of equations for the six osculating orbit elements of each orbit.  In a selected basis of coordinates, the elements themselves can be derived algebraically from $\bm x$ and $\bm v$ (see Sec.\ 3.3 of \cite{PW2014}, for example).  For a purely Keplerian two-body orbit, the elements are constant.

A hierarchical triple is one in which the inner binary and the third body are sufficiently well separated that the added gravitational fields induce small perturbations on the Keplerian motion of the inner and outer orbits (see Sec.\ \ref{sec:secevol} for precise statements of the conditions).  We expect that the perturbations will induce small variations of the orbit elements that occur on two basic timescales: the two orbital periods, and a longer, perturbative timescale associated with the slow secular evolution of the elements (the advance of the perihelion of Mercury being a leading example of the latter).  To treat these two kinds of evolution, we adopt the ``two-timescale'' approach from applied mathematics \cite{1978amms.book.....B}.  In our implementation, we split each orbit element into an ``average part'', that is, a part averaged over both orbital periods, and a perturbative, oscillatory ``average-free'' part, and obtain an equation for the long-term evolution of the average part of each element, and an equation for the average-free part.  This method naturally lends itself to going to higher orders in perturbation theory, basically by plugging the solution (average plus average-free part) for each orbit element back into the LPE and repeating the procedure.

In Paper I \cite{2021PhRvD.103f3003W}, we carried out this procedure and obtained the leading second-order ``quadrupole-quadrupole'' contributions to the evolution of the average orbit elements of the inner orbit.  We studied the astrophysical consequences of these effects, particularly for systems where the third mass is much larger than the binary mass. High outer mass systems are of interest because the quadrupole perturbations for inner orbit elements have amplitudes proportional to $(m_3/m)\epsilon^3$, where $m_3$ and $m$ are the masses of the outer body and inner binary, respectively, and $\epsilon = a/A$, where $a$ and $A$ are the inner and outer semimajor axes, and therefore quadrupole-squared perturbations should have amplitudes $(m_3/m)^2 \epsilon^6$ (for technical reasons related to the orbit averaging process, the actual amplitudes were larger, proportional to  $(m_3/m)^{2} (1+m_3/m)^{-1/2} \epsilon^{9/2}$). In contrast, the sequence of first-order perturbations at various orders in a multipole expansion are proportional to $m_3/m$.  As a result, we found negligible effects of quadrupole-squared terms on systems with low-mass third bodies, such as ``hot-Jupiters'' orbiting a solar mass star, perturbed by a distant low-mass brown dwarf \cite{2011Natur.473..187N,2013MNRAS.431.2155N}, while for systems such as a stellar-mass binary orbiting a supermassive black hole, the quadrupole-squared terms suppressed orbital flips that were induced by first-order perturbations at octupole order (Fig.\ 5 of Paper I). 

In Paper II \cite{2010PhRvD..82d4028C}, we completed the procedure, obtaining the first-order perturbative terms through ``dotriacontapole'' order in a multipole expansion of the perturbing fields, corresponding to effects at order $(m_3/m)\epsilon^6$, along with an array of second-order effects with dependences $\epsilon^{9/2}$, $\epsilon^5$, $\epsilon^{11/2}$ and $\epsilon^6$.  While in Paper I, we focused only on the inner orbit, in Paper II, we obtained equations for both the inner and outer orbit elements.  For simplicity, we denote these second-order and dotriacontapole effects by the inelegant acronym SOD.  We have made available the complete set of evolution equations for inner and outer orbits, including all first-order equations from quadrupole through dotriacontapole order, and all second-order equations as described above, at the public site  \url{github.com/landenconway/Three-Body-Secular-Equations}.

In this paper, we explore some of the astrophysical consequences of the SOD terms.  For some cases we also compare our results including SOD effects with direct numerical evolution of the three-body equations of motion. Here we summarize some of the main conclusions. 

\begin{itemize}
\item 
For systems with low-mass outer bodies, such as the ``hot Jupiter'' example, the SOD effects on the inner orbit are small, as expected, making only minor shifts in times between orbital flips (see Figs.\ 4, 5 and 6).
\item 
For some high outer-mass systems, such as binaries orbiting supermassive black holes, SOD effects do not alter the behavior induced by the leading quadrupole-squared terms discussed in Paper I, which cleanly suppressed orbital flips when various perturbative timescales are well separated, producing more ragged flip behavior otherwise.  The results are supported by direct N-body simulations (see Fig.\ 7).
\item 
In Paper II, we found that second-order terms introduced variations in the semimajor axes, beginning at the level of $\epsilon^5$.  We found such variations to occur in the semimajor axis of a planet orbiting a 10:1 mass ratio binary, in agreement with N-body simulations (see Figs.\ 2 and 8).  We suggest that this class of systems -- circumbinary planets -- may be particularly susceptible to astrophysically interesting variations of the planet's semimajor axis. 
\item 
In an example of a hierarchical triple of three black holes without orbital flips, we found that SOD effects induced more extreme eccentricity in the inner binary than did the octupole-level equations.  Because gravitational radiation reaction is extremely sensitive to high eccentricities, this could have important implications for the time to merger of the inner binary.  We suggest that simulations of populations of such triples that seek to predict inspiral event rates for gravitational-wave detection should use the full array of SOD equations, rather than simple first-order equations, which could yield biased results.  Integrating the SOD equations should be computationally less burdensome than direct integrations of the equations of motion.
\end{itemize}

The remainder of the paper provides details.  In Sec.\ \ref{sec:triples} we briefly review the basic method and results of Paper II.  In Sec.\ \ref{sec:astro} we consider specific cases, some studied in earlier papers \cite{2017PhRvD..96b3017W,2021PhRvD.103f3003W}, some new.  Some case were chosen with view to illustrating the variations in the averaged semimajor axes predicted by second-order terms.  In Sec.\ \ref{sec:discussion}, we make concluding remarks.

\section{Secular evolution of hierarchical triples}
\label{sec:triples}

\subsection{Lagrange planetary equations}

\begin{figure}[t]
\begin{center}

\includegraphics[width=3.2in]{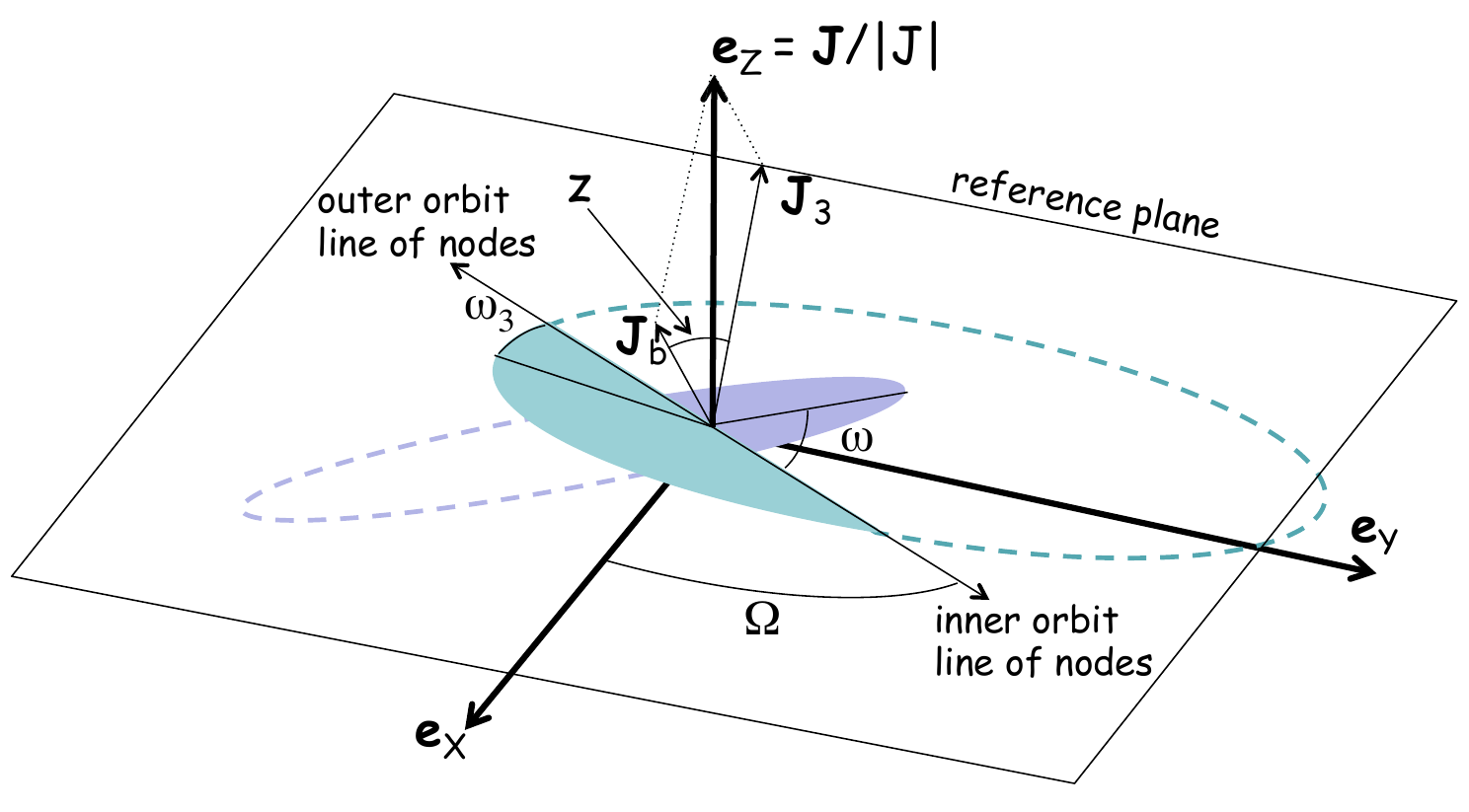}

\caption{Orientation of inner and outer orbits. (Color figures in online version.)
\label{fig:orbits} }
\end{center}
\end{figure}

We consider a hierarchical three-body system illustrated in Fig.\ \ref{fig:orbits}, with bodies 1 and 2 comprising the ``inner'' binary, and with body 3 taken to be the ``outer'' perturbing body.  The orbital separation of the inner binary is assumed to be small compared to that of the outer binary.  We define $m \equiv m_1 + m_2$, $M \equiv m + m_3$, $\eta \equiv m_1m_2/m^2$ with the convention that $m_1 \le m_2$, and $\eta_3 \equiv m_3 m/M^2$ .   To the leading order
in the ratio of $r$ to $R$, where $r \equiv |{\bm x}| = |{\bm x}_1 - {\bm x}_2|$ is the inner binary separation, and $R \equiv |{\bm X}| = |{\bm x}_3 - {\bm x}_{\rm cm}|$ is the separation between the outer body and the center of mass of the inner binary, known as ``quadrupole'' order, the equations of motion take the form
\begin{align}
a^j &= - \frac{Gm n^j}{r^2} + \frac{Gm_3 r}{R^3} \left ( 3 N^j N_n - n^j \right ) 
 \,,
\nonumber \\
A^j &=-  \frac{GM N^j}{R^2} - \frac{3}{2} \frac{GM\eta r^2}{R^4}  \left ( 5 N^j N_n^2 - 2 n^j N_n - N^j  \right )
 \,,
\label{eq2:eom3}
\end{align}
where $\bm{a} \equiv d^2\bx/dt^2$, $\bm{A} \equiv d^2\bm{X}/dt^2$, $\bm{n} \equiv \bx/r$, $\bm{N} \equiv \bm{X}/R$, $N_n \equiv \bm{N} \cdot \bm{n}$, and $G$ is Newton's constant. An expression valid to all multipole orders is given in Eq.\ (1) of Paper I.  

We define the osculating orbit elements of the inner and outer orbits in the standard manner: for the inner orbit, we have the orbit elements $p$, $e$, $\omega$, $\Omega$ and $\iota$, with the definitions
\begin{eqnarray}
r &\equiv& p/(1+e \cos f) \,,
\nonumber \\
{\bm x} &\equiv& r {\bm n} \,,
\nonumber \\
{\bm n} &\equiv& \left [ \cos \Omega \cos(\omega + f) - \cos \iota \sin \Omega \sin (\omega + f) \right ] {\bm e}_X 
\nonumber \\
&&
 + \left [ \sin \Omega \cos (\omega + f) + \cos \iota \cos \Omega \sin(\omega + f) \right ]{\bm e}_Y
\nonumber \\
&&
+ \sin \iota \sin(\omega + f) {\bm e}_Z \,,
\nonumber \\
{\bm \lambda} &\equiv& d{\bm n}/df \,, \quad \hat{\bm h}={\bm n} \times {\bm \lambda} \,,
\nonumber \\
{\bm h} &\equiv& {\bm x} \times {\bm v} \equiv \sqrt{Gmp} \, \bm{\hat{h}} \,,
\label{eq2:keplerorbit1}
\end{eqnarray}
where (${\bm e}_X,\,{\bm e}_Y ,\,{\bm e}_Z$) define a reference basis, with ${\bm e}_Z$ aligned along the total angular momentum of the system, and with the ascending node of the inner orbit oriented at an angle $\Omega$ from the $X$-axis.  From the given definitions, we infer that ${\bm v} = \dot{r} {\bm n} + (h/r) {\bm \lambda}$ and $\dot{r} = (he/p) \sin f$. 

The outer orbit is defined in the same manner, with orbit elements $P$, $E$, $\omega_3$, $\Omega_3$, and $\iota_3$ replacing $p$, $e$, $\omega$, $\Omega$ and $\iota$, $\bm{\Lambda}$ and $\bm{H}$ replacing $\bm{\lambda}$ and $\bm{h}$, and $F$ replacing $f$.  
The semimajor axes of the two orbits are defined by $a \equiv p/(1-e^2)$ and $A \equiv P/(1-E^2)$.

With the orbits and basis defined this way, it is straightforward to show that
\begin{eqnarray}
\Omega_3 &=& \Omega + \pi \,,
\quad
J_b \sin \iota =  J_3 \sin \iota_3 \,,
\end{eqnarray}
where $J_b =  m \eta \sqrt{Gmp}$ and $J_3 = M\eta_3 \sqrt{GMP}$, and, using the exact Lagrange planetary equations, that these relations hold for all time.
Defining
\begin{align}
\beta &\equiv \frac{J_b}{J_3} = \frac{\sin \iota_3}{\sin \iota} \,,
\qquad
z \equiv \iota + \iota_3 \,,
\label{eq2:betaz}
\end{align}
it is straightforward to obtain the relations
\begin{equation}
\cot \iota = \frac{\beta + \cos z}{\sin z} \,,  \qquad \cot \iota_3 = \frac{\beta^{-1} + \cos z}{\sin z} \,,
\label{eq2:inclinations}
\end{equation}
so that only the {\em relative} inclination $z$ between the two orbits is dynamically relevant; given an evolution for $z$ and $\beta$, the individual orbital inclinations can be recovered algebraically from Eqs.\ (\ref{eq2:inclinations}).  For most hierarchical systems of interest, $\beta$ is small, but it can be large, particularly for systems with very low-mass third bodies.

We define the perturbing accelerations $\delta {\bm a} \equiv \bm{a} + Gm{\bm n}/r^2$ and $\delta {\bm A} \equiv \bm{A} + GM{\bm N}/R^2$,
and from them the radial, cross-track and out-of-plane components 
${\cal R} \equiv {\bm n} \cdot \delta {\bm a}$,
 ${\cal S} \equiv {\bm \lambda} \cdot \delta {\bm a}$ and
 ${\cal W} \equiv \bm{\hat{h}} \cdot \delta {\bm a}$, together with their counterparts for the outer orbit,
and we write down the Lagrange planetary equations for the evolution of the orbit elements,
\begin{eqnarray}
\frac{dp}{dt} &=& 2 \sqrt{\frac{p^3}{Gm}}\, \frac{{\cal S}}{1+e \cos f} \,,
\nonumber \\
\frac{de}{dt} &=& \sqrt{\frac{p}{Gm}} \left [ \sin f \, {\cal R} + \frac{2\cos f + e +e\cos^2 f}{1+ e\cos f} {\cal S} \right ]\,,
\nonumber \\
\frac{d\varpi}{dt} &=& \frac{1}{e}\sqrt{\frac{p}{Gm}} \left [ -\cos f \, {\cal R} + \frac{2 + e\cos f}{1+ e\cos f}\sin f {\cal S} 
\right ] \,,
\nonumber \\
\frac{d\iota}{dt} &=& \sqrt{\frac{p}{Gm}}\, \frac{\cos (\omega +f)}{1+ e\cos f} {\cal W} \,,
\nonumber \\
 \frac{d\Omega}{dt} &=& \sqrt{\frac{p}{Gm}} \,\frac{\sin (\omega +f)}{1+ e\cos f} \frac{\cal W}{\sin \iota}\,.
\label{eq2:lagrange}
\end{eqnarray}
The auxiliary variable $\varpi$ is defined such that the change in pericenter angle is given by $\dot{\omega} = \dot{\varpi} -  \dot{\Omega} \cos \iota$.  These are augmented by a ``sixth'' planetary equation that relates $f$ to time via
\begin{align}
\frac{df}{dt} 
&= \left (\frac{df}{dt} \right )_K - \frac{d\varpi}{dt}\,,
\label{eq:dfdt}
\end{align}
where $(df/dt)_K = (Gm/p^3)^{1/2} (1+e \cos f)^2$ denotes the Keplerian expression.    
The LPE for the outer binary take the form of Eqs.\ (\ref{eq2:lagrange}) and (\ref{eq:dfdt}), with suitable replacements of all the relevant variables.

\subsection{Secular evolution to dotriacontapole order}
\label{sec:secevol}

From Eqs.\ (\ref{eq2:eom3}), we can identify the leading, quadrupolar perturbative parameter for the inner orbit as $\alpha \epsilon^3$, where $\alpha \equiv m_3/m$ and $\epsilon \equiv a/A$, and for the outer orbit (with time in units of one inner orbit period) as $\alpha \epsilon^3 \beta$.  The other key parameter is $\epsilon$ itself, which controls the effects of higher multipoles.
In Paper II, using a two timescale method, we obtained a perturbative solution to the long term evolution of the averaged orbital elements of hierarchical triple systems that is complete through second order in the perturbative parameter and through $\epsilon^6$ (dotriacontapole) in the multipolar 
expansion. We obtained equations for the evolution of both the inner and outer orbits. We placed no formal restrictions on the masses, orientations, or inclinations of the orbits beside requiring that the perturbative parameter $\alpha \epsilon^3 \ll 1$ and that $\epsilon \ll 1$. Additionally, we imposed no formal restrictions on the orbital eccentricities $e$ and $E$. However, resonant interactions between the binaries grow as the outer $E$ increases. The ``secular approximation'' that we used to average over the two orbital periods ignores orbital resonance effects and will begin to fail in describing systems with high $E$. In practice, we will confine our attention to systems that satisfy the improved Mardling-Aarseth stability criterion \cite{2001MNRAS.321..398M,2022MNRAS.516.4146V}; roughly speaking, this phenomenological criteron separates systems where resonant interactions induce large changes in the semimajor axes, from those that do not experience such instability inducing effects.

\begin{table}[t]
\caption{List of perturbative effects}
\medskip
\begin{tabular}{l@{\hskip 1 cm}l@{\hskip 0.5cm}c@{\hskip 0.5cm}c}
\hline
\vspace{-0.3cm}\\
%Column headings here
Contribution&Amplitude&$da/d\tau$&$dA/d\tau$\\
\vspace{-0.3cm}\\
\hline  
\vspace{-0.3cm}\\
%Column contents here
\multicolumn{3}{l}{\em First-order perturbations}\\
Quadrupole&{$\alpha \epsilon^3$}&0&0\\
Octupole&{$\alpha\Delta\epsilon^4$}&0&0\\
Hexadecapole&{$\alpha(1-3\eta)\epsilon^5$}&0&0\\
Dotriacontapole&{$\alpha\Delta(1-2\eta)\epsilon^6$}&0&0\\
\vspace{-0.3cm}\\
\hline
\vspace{-0.3cm}\\
\multicolumn{3}{l}{\em Second-order feedback}\\
$Q_{{\rm in,in}}^{\rm quad} \int^t Q^{\rm quad}_{\rm in}$&{$\frac{\alpha^2}{\sqrt{(1+\alpha)}} \epsilon^{9/2}$}&$0$&0\\
$Q_{{\rm in,out}}^{\rm quad} \int^t Q^{\rm quad}_{\rm out}$&{${\alpha}\eta \epsilon^{5}$}&$0$&$\ne0$\\
$Q_{{\rm in,in}}^{\rm (quad} \int^t Q^{\rm oct)}_{\rm in}$&{$\frac{\alpha^2\Delta}{\sqrt{(1+\alpha)}}  \epsilon^{11/2}   $}&$0$&0\\
$Q_{{\rm in,out}}^{\rm (quad} \int^t Q^{\rm oct)}_{\rm out}$&{$\alpha\eta\Delta\epsilon^{6}$}&$0$&$\ne0$\\
$Q_{{\rm in,in}}^{\rm quad} \int^t Q^{\rm quad}_{\rm in}$&{$\alpha^2 \epsilon^6$}&$\ne 0$&$\ne0$\\

\vspace{-0.3cm}\\
\hline
\vspace{-0.3cm}\\
\multicolumn{3}{l}{\em Second-order time conversion}\\
$Q_{\rm in}^{\rm quad}  (dt/dF)^{\rm quad}$&{$\alpha \eta \epsilon^{5}$}&$\ne 0$&$\ne0$\\
$Q_{\rm in}^{\rm quad}  (dt/df)^{\rm quad}$&{$\alpha^2 \epsilon^6$}&$\ne 0$&$\ne0$\\
$Q_{\rm in}^{\rm (quad}  (dt/dF)^{\rm oct)}$&{$\alpha \eta \Delta \epsilon^{6}$}&$\ne 0$&$\ne0$\\
\vspace{-0.3cm}\\
\hline
\end{tabular}
\label{tab:table1}
\end{table}

At our working order, there are twelve unique effects that contribute to the overall secular evolution of a given orbital element. Table \ref{tab:table1} lists these twelve effects and their respective amplitudes for an inner orbit element. The outer orbit's amplitudes are found by multiplying those of the inner orbit by $\beta$. Note that $\Delta =(m_1 - m_2)/m$.

The $Q$'s in Table \ref{tab:table1} refer to a multipolar contribution to the Lagrange planetary equation for a given element $X_\alpha$, i.e. $dX_\alpha/dt = Q_\alpha (X_\beta, t)$.  ``First-order perturbations'' refer to averaging the planetary equations over the two orbits holding the orbit elements fixed. These are the basic multipolar effects, from quadrupole to dotriacontapole. ``Second-order feedback'' refers to feeding the periodic, average-free solutions (represented schematically by $\int^t Q_\beta dt$) for the inner and outer elements back into the expression $Q_\alpha$ for a given element, expanding in powers of the perturbation parameter, and orbit-averaging again (see Paper I, Sec.\ III.C for discussion of the subtleties involved in averaging integrals of functions). The notation ``(quad $\dots$ oct)'' (the parentheses denote symmetrization) implies that octupole variations are fed back into quadrupole perturbations and vice versa. ``Second-order time conversion'' refers to using Eq.\ (\ref{eq:dfdt}) and its analogue for the outer orbit to convert time averages to averages over the orbital phases.  We truncated our examination of effects at order $\epsilon^6$. Also shown in the table is whether the semimajor axes are predicted to be constant under the perturbation.  See Paper II for detailed discussion of these contributions and for formulae for almost all of them. See
\url{github.com/landenconway/Three-Body-Secular-Equations} for the complete set of expressions.

\subsection{Non-constancy of the semi-major axes}
\label{sec:smavariations}

An unexpected result of our calculations in Paper II was the presence of secular variations in the averaged 
semimajor axes $a$ and $A$. At linear order in perturbation theory, $da/dt = dA/dt = 0$ at all multipole orders, and we provided a simple proof of this in Appendix D of Paper II. However, at second order in perturbation theory we found this no longer to be true. Secular variations in the axes first arise at the equivalent of hexadecapole order $\epsilon^5$ from two origins; the feedback of quadrupole perturbations of the outer orbital elements into the quadrupole perturbations and corrections to the relation between the outer true anomaly and time. These variations have the form
%\begin{widetext}
\begin{align}
\frac{da}{d\tau} &=\frac{3 \pi a}{128} \frac{\alpha \eta \epsilon^5 (10 +E^2)(1-e^2)^{1/2}}{(1-E^2)^{7/2}}  
\nonumber \\
& \quad
  \times \biggl [ 6 (2+3e^2) \cos z \sin^2 z \sin 2\omega_3 
  \nonumber \\
  & \quad + 5e^2 \left [ (1 \mp \cos z)^2 
  (2 \pm 3 \cos z) \sin (2\omega \pm 2\omega_3 ) 
  \right ] \biggr ] \,,
  \nonumber \\
  \frac{dA}{d\tau} &= - \frac{9\pi A}{16384} \frac{\alpha \eta \beta \epsilon^5 }{(1-e^2)^{1/2}(1-E^2)^{9/2}} 
  \nonumber \\
  &\quad 
  \times \biggl [ 2 \sin^2 z \biggl \{ {\cal G}_{1} \sin 2\omega_3 + E^2(4-E^2) 
  \nonumber \\
  & \qquad\quad
  (8+72e^2-41e^4) \sin^2 z \sin 4 \omega_3 \biggr \}
\nonumber \\ 
& \qquad 
+(64+16E^2-E^4)  \biggl \{4[\pm e^2  {\cal G}^\pm_{2} \sin(2\omega \pm 2\omega_3)]
\nonumber \\
& \qquad \quad
+39 \sin^2 z[\pm e^4  (1 \mp \cos z)^2 \sin(4\omega \pm 2\omega_3)] \biggr \}
\nonumber \\ 
& \qquad
+ E^2 (4-E^2) \biggl \{ 4(22-9e^2) \sin^2 z 
\nonumber \\
&\qquad \quad 
[\pm e^2 (1 \mp \cos z)^2 \sin(2\omega \pm 4\omega_3)]
\nonumber \\
&\qquad \quad 
 +13[ \pm e^4 (1\mp \cos z)^4 \sin(4\omega \pm 4\omega_3)] \biggr \} \biggr ] \,,
 \label{eq:dadtdAdt}
\end{align}
where $\tau$ is time in units of the inner orbital period, i.e., $\tau = (t/2 \pi)(Gm/a^3)^{1/2}$.
We adopt a notation whereby $[A^\pm]= A^+ + A^-$,
$[(a \mp b) B^\pm] = (a-b)B^+ + (a+b)B^-$, and so on, and where
\begin{align}
{\cal G}_{1} & = (64+16E^2-E^4) \left (3(8+72e^2-41e^4)\cos^2 z
\right .\nonumber \\
& \qquad
\left .
+ (24+88e^2-99e^4) \right ) \,,
\nonumber \\
{\cal G}^\pm_{2} &=  (22-9e^2) (1\mp \cos z)^2 (1 \pm 3\cos z +3\cos^2 z) \,.
\end{align}

In Paper II we argued that these variations do not contradict famous theorems by  Poisson, Poincar\'e, Tisserand and others \cite{1897BuAsI..14..241P,Tisserand89,1902AnPar..23A...1A,1978A&A....68..199D}.
Those theorems prove only that the semimajor axes contain no ``purely secular'' terms, i.e. terms proportional to $t$.  The variations shown in Eqs.\ (\ref{eq:dadtdAdt}) are actually {\em periodic} in linear combinations of the two pericenter angles, fully consistent with the theorems.

With these unexpected results, we particularly  want to test our secular equations against N-body numerical calculations, paying special attention to the semimajor axes. What we found is that making this comparison is not simple.

\begin{figure}[t]
\begin{center}

\includegraphics[width=3.2in]{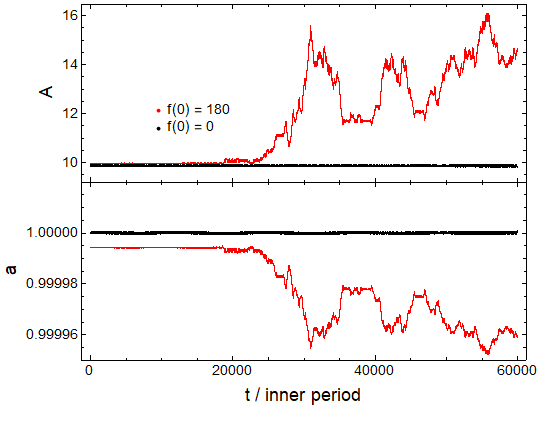}

\caption{Semimajor axis evolution in two N-body numerical calculations of a circumbinary system, differing only in the initial inner true anomaly. The black and red curves correspond to initial $f = 0$ and  $f = 180^{\rm o}$, respectively, while $F=0$ in both cases. The other orbit elements do not show such differences.  The parameters of the system are $m_1 = M_\odot$, $m_2 = 10M_\odot$, $m_3 = M_\odot/1000$, $a = 1$, $A = 10$, $e = 0.25$, $E = 0.6$, $\omega = \omega_3 = 0^{\rm o}$, and  $z = 65^{\rm o}$.}
\label{fig:axesproblem1} 
\end{center}
\end{figure}

\begin{figure}[t]
\begin{center}

\includegraphics[width=3.2in]{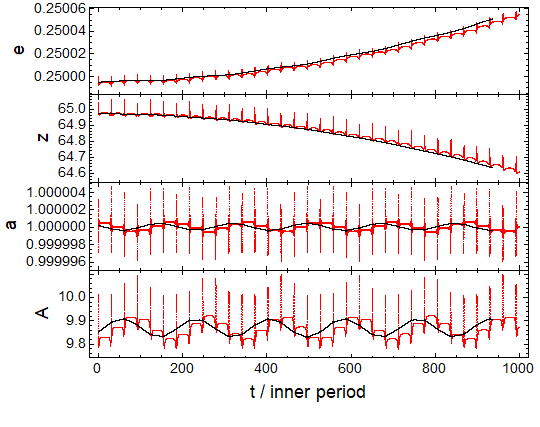}

\caption{ Comparison of the four orbital elements $e$, $z$, $a$ and $A$ in an N-body numerical calculation (red). The secular change (black) in $e$ and $z$ over the 1000 orbits is much larger than their respective short timescale variations. This is in contrast to $a$ and $A$ whose secular changes are small compared to their short timescale variation. The parameters describing this system are the same as in Fig.\ \ref{fig:axesproblem1}.}

\label{fig:axesproblem2} 
\end{center}
\end{figure}

First, in a numerical N-body code, one must specify the initial orbital phases of the inner and outer binary, usually in the form of the true anomalies $f$ and $F$. By contrast, the secular equations are independent of the orbital phases because they have been averaged away. For most systems, the secular evolutions of the orbital elements $e, E, \omega, \omega_3$ and $z$ remain relatively consistent between variations in the initial orbital phases, whereas the evolutions of the semimajor axes seem to be more sensitive to the initial phases. An example of this is shown in Fig.\ \ref{fig:axesproblem1}, for a planet orbiting an asymmetric binary system.  
The N-body results were obtained using the REBOUND code \cite{rebound}, integrated using IAS15, a 15th order Gauss-Radau integrator. 
We note here that circumbinary planetary systems of a type described in Fig.\ \ref{fig:axesproblem1} may be particularly susceptible to SOD induced variations in semimajor axes; for another example, see Sec.\ \ref{sec:circumbinary}.

Secondly, the secular equations are based on the averaged orbital elements,  while an N-body calculation uses the instantaneous osculating orbital elements as initial conditions. In our formulation, these are not the same, but are related by a perturbative, average-free part (see Paper II Eq.\ (19)  for details). 
%Therefore, and N-body simulation initiated using values of the orbit elements  a state characterized by a set of values for its orbit-averaged elements will be slightly different than a state defined by those values used as initial instantaneous orbit elements.
One could calculate the corresponding initial values of the average-free pieces to some order in the perturbation expansion, in order to allow for more accurate mapping between the averaged and instantaneous elements.  Given the extreme sensitivity of three-body dynamics to initial conditions, there is no guarantee that this would be effective, so we have not tried this. Instead, we will select a set of values for the orbit elements as initial values for integrating our SOD equations and the {\em same} values as initial conditions for the N-body integrations. 

In comparing N-body calculations with other results, we carry out a running average of the N-body data with a window that includes enough orbital timescales to effectively average out orbital variations, without averaging away variations on perturbative timescales.  The window size used depends on the parameters of each system.  One result of this is that there will occasionally appear an offset between SOD and N-body curves of a given orbit element.

Third, all the orbit elements experience quadrupole-order fluctuations on orbital timescales.  Many of them experience secular variations that ultimately dominate over the short-term variations and are easily identified in a numerical simulation.  Examples are the Kozai-Lidov variations in eccentricity and inclination and the secular variations in the pericenter and nodal angles. But the semimajor axes average almost perfectly to zero, making it difficult to measure in the numerical data the small or long-term changes induced by second-order perturbations.
%A further issue with the semi major axes comes from their behavior on the short orbital timescale. The axes will often oscillate on the orbital timescale with an amplitude comparable or larger to the amplitude of the secular oscillations. This in turn makes it difficult to identify the secular  changes among the larger orbital fluctuations. 
An example of this is illustrated in Fig.\ \ref{fig:axesproblem2}, for the same circumbinary system plotted in Fig.\ \ref{fig:axesproblem1}, where the orbital fluctuations in $a$ and $A$ are much larger than their secular changes, in contrast to the behavior of $e$ and $z$.  
%elements as the orbital fluctuations are much smaller than the secular oscillations, although the outer eccentricity can sometimes behave more similarly to the axes. This is also evident from the fact that the semi major axes secular variations first appear analytically at second order, making them small by default.

\section{Astrophysical implications}
\label{sec:astro}

\subsection{Revisiting previous systems}  
In this section we revisit a number of systems that were previously studied in \cite{2017PhRvD..96b3017W} using linear hexadecapole order effects and in Paper I using only the top feedback effect in the middle section of Table \ref{tab:table1}, denoted the ``dominant quadrupole-squared term''. We compare the previous secular evolutions to the newly generated evolutions using the full set of SOD terms from Paper II. In comparing the two, the evolution of the semimajor axes will always differ since linear effects predict zero secular variation while some second-order effects predict non-zero variations. However, in most cases the predicted secular variations in the semimajor axes are very small.

Three-body systems are prone to unstable behavior induced by orbital resonances, particularly involving higher temporal harmonics of the perturbations caused by outer bodies on eccentric orbits.  To avoid these cases, we focus on systems that are deemed stable according to the  
 criterion from \cite{2022MNRAS.516.4146V}, which is an improvement on the widely used stability criterion given by Mardling and Aarseth \cite{2001MNRAS.321..398M}.  Using our notation, we define a quantity
 \begin{align}
 Y^* &\equiv \frac{5}{12} \epsilon^{-1} \left ( \frac{m}{M} \right )^{2/5} 
 \left (\frac{(1-E)^2}{(1+\tilde{e})} \right )^{3/5}
 \nonumber \\
 & \quad \times \left [ 1 - \frac{1}{4}(1-0.2 \tilde{e} +E) \sin^2 \frac{z}{2} \right ]^{-1} \,,
 \label{eq:Ystar}
 \end{align}
where 
\begin{equation}
\tilde{e} \equiv {\rm max}\left [ e, \tfrac{1}{2} \left (1- \tfrac{5}{3} \cos^2 z \right ) \right ] \,.
\end{equation}
 For each system considered, we provide the stability ratio $Y^*$ calculated using Eq.\ (\ref{eq:Ystar}),  where ratios larger than one imply stability and ratios less than one imply instability. 

\subsubsection{Hot Jupiters}
The first system we consider is taken from Naoz et al.\ \cite{2011Natur.473..187N} who discussed the possibility of hot Jupiters in retrograde orbits as a result of secular three-body effects. They considered a system composed of an inner binary containing a solar mass star and a Jupiter mass planet ($M_\odot = 1047 M_J$) with  $a = 6$ a.u.\,  perturbed by a brown dwarf of mass $40 M_J$ and $A = 100$ a.u. The remaining initial orbital elements are 
\begin{equation}
e = 0.001,\, \omega = 45^{\rm o},\, E = 0.6,\, \omega_3 = 0^{\rm o},\, z = 65^{\rm o}. \nonumber
\end{equation}

\begin{figure}[t]
\begin{center}

\includegraphics[width=3.2in]{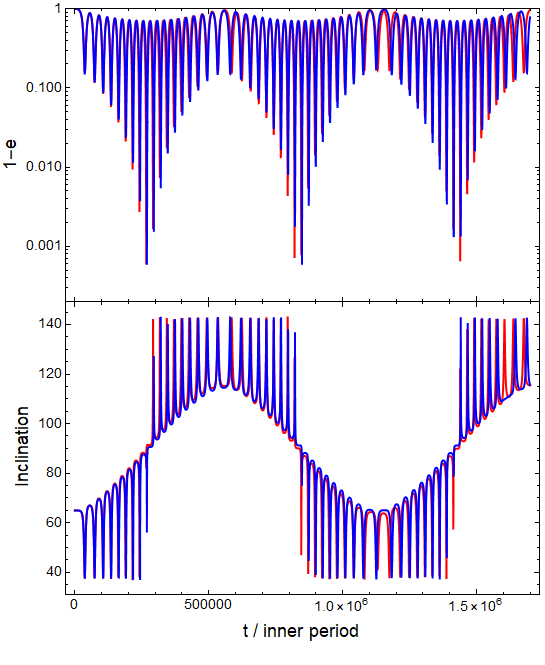}

\caption{ Comparison of orbital flips and eccentricity excursions in a Sun-Jupiter-Brown-Dwarf system between linear hexadecapole effects (blue) and SOD effects (red).}

\label{fig:HJ} 
\end{center}
\end{figure}

The stability factor  $Y^*$ of this system is 2.138. We evolve the secular equations over $1.7 \times 10^6$ periods of the inner orbit's initial Keplerian period.  The differences in the secular evolution between the hexadecapole (blue) and SOD (red) results are minimal as represented in the inclination and eccentricity evolutions in Fig.\ \ref{fig:HJ}. This is not a surprise, as already at octupole order the secular equations are in good agreement with exact N-body calculations \cite{2011Natur.473..187N}.

\subsubsection{Orbital flips from nearly coplanar orbits}

\begin{figure}[t]
\begin{center}

\includegraphics[width=3.2in]{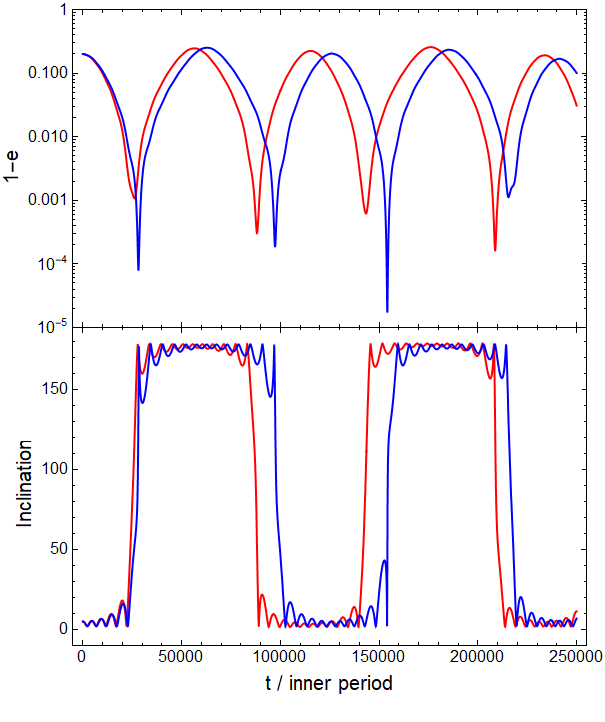}

\caption{ Comparison of orbital flips and eccentricity excursions in a nearly coplanar Jupiter Sun system between linear hexadecapole effects (blue) and SODeffects (red).}

\label{fig:CP} 
\end{center}
\end{figure}

The possibility of orbital flips from nearly coplanar orbits was discovered by Li et al.\ \cite{2014ApJ...785..116L} in a similar system of masses but with $a = 4$ a.u.\ and $A = 50$ a.u. The remaining initial orbital elements are
\begin{equation}
e = 0.8,\, \omega = 0^{\rm o},\, E = 0.6,\, \omega_3 = 0^{\rm o},\, z = 5^{\rm o}. \nonumber
\end{equation}
The stability factor  $Y^*$ of this system is 1.202. The system is evolved for $2.5 \times 10^5$ periods of the inner orbit's initial Keplerian period. The results are shown in Fig.\ \ref{fig:CP}. The inclusion of SOD effects causes two main differences. The timescale over which the orbits remain prograde or retrograde has shortened and the peaks of the eccentricity excursions are smaller and less varied. 

\subsubsection{A triple-star hierarchical system}

\begin{figure}[t]
\begin{center}

\includegraphics[width=3.2in]{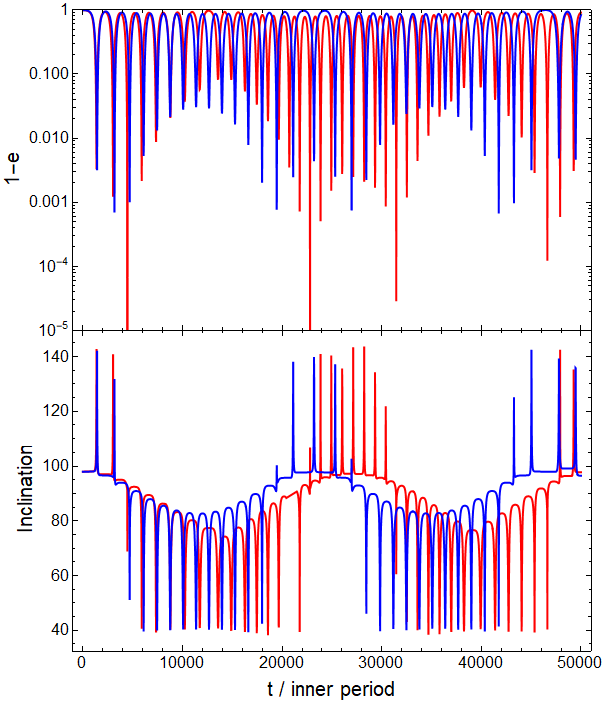}

\caption{ Comparison of orbital flips and eccentricity excursions in a hierarchical triple-star system between linear hexadecapole effects (blue) and SOD effects (red).}

\label{fig:TS} 
\end{center}
\end{figure}

This system, studied by Fabrycky and Tremaine \cite{2007ApJ...669.1298F}, consists of an inner binary with $m_1 = M_\odot$,  $m_2 = 0.25 M_\odot$, $a = 60$ a.u. and an  outer star, with $m_3 = 0.6 M_\odot$, $A = 800$ a.u. The remaining initial orbital elements are
\begin{equation}
e = 0.01,\, \omega = 0^{\rm o},\, E = 0.6,\, \omega_3 = 0^{\rm o},\, z = 98^{\rm o}. 
\nonumber
\end{equation}
The stability factor  $Y^*$ of this system is 1.588. We evolve the system for $5 \times 10^4$ periods of the inner orbit's initial Keplerian period. The results are shown in Fig.\ \ref{fig:TS}. The differences amount to a small increase in the length of time for prograde motion compared to linear hexadecapole order, slightly more regular oscillations during retrograde motion, and larger 
 predicted eccentricity values during the orbital flips. The quantity $1-e$ is over 100 times smaller at the orbital flip than for linear hexadecapole motion.
% The discrepancy in the eccentricities is most notable during orbital flips with dotriacontapole solutions predicting eccentricities up to 100 times larger than the octupole solution when comparing $1-e$. 
 This amounts to predicting a pericenter distance that is over 100 times smaller, a nontrivial amount that could have significant astrophysical 
 implications.
 
\subsubsection{ High-outer-mass systems}

\begin{figure}[t]
\begin{center}

\includegraphics[width=3.2in]{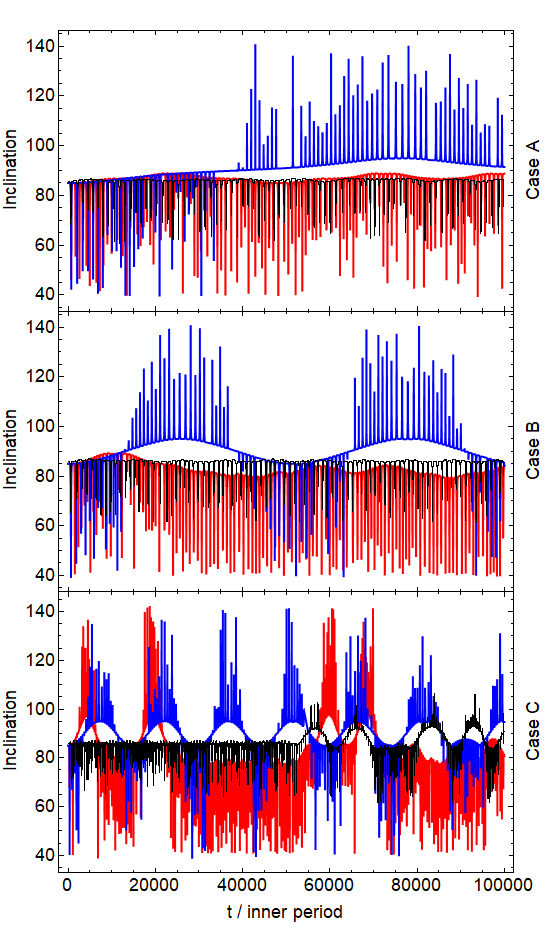}

\caption{ Suppression of orbital flips in high-outer-mass triple systems. The octupole, SOD and N-body solutions are represented by blue, red, and black, respectively. The inclusion of SOD effects suppresses orbital flips in Cases A and B, in agreement with the N-body solution. }

\label{fig:CaseABC} 
\end{center}
\end{figure}

Because the second-order perturbative effects involve an additional factor of $m_3/m$, then the regime of high outer mass, where $m_3/m \gg 1$, may display interesting effects.  
In Paper I, we studied systems such as a $1 M_\odot + 100 M_\odot$ binary system orbiting a $10^6 \, M_\odot$ black hole, and found that the dominant quadrupole squared terms (the leading second-order feedback terms in Table \ref{tab:table1}) suppressed the orbital flips that were generated by purely first-order effects, mainly at octupole order.   We continue that study here by looking at three similar systems to see if the addition of SOD terms will cause the suppression of orbital flips to persist. We consider the same inner binary, with an initial semimajor axis $a = 0.01$ a.u.    We consider three different  cases A, B and C for the outer binary, with  $m_3 = 10^6 \,M_\odot$, $A = 3$ a.u., $m_3 = 10^5 \,M_\odot$, $A = 1.35$ a.u. , and $m_3 = 10^4M_\odot$, $A = 0.5$ a.u.,  respectively.  It turns out that the original cases A, B, and C studied in Paper I are unstable according to Eq.\ (\ref{eq:Ystar}), and so we have slightly increased the initial outer semimajor axis of each case to ensure stable evolution. 
The remaining initial orbital elements are
\begin{equation}
e = 0.01,\, \omega = 0^{\rm o},\, E = 0.6,\, \omega_3 = 0^{\rm o},\, z = 85^{\rm o}. \nonumber
\end{equation}
The stability factors $Y^*$ for the three cases are 0.9957, 1.125 and 1.043. We evolve the systems for $10^5$ periods of the inner orbit's initial Keplerian period. The results are shown in Fig.\ \ref{fig:CaseABC} with the octupole only, SOD, and N-body evolutions represented by blue, red, and black, respectively. 
 
%The N-body plot points were sampled evenly over the integration time and when the outer binary was not at pericenter (this will be true for all N-body simulations shown). 
In cases A and B, where second-order effects have a larger amplitude than octupole effects, the orbital flips that are present in the octupole solution are cleanly suppressed by SOD effects, consistent with the N-body solution. In case C, the octupole effects are comparable to the largest second-order effect, leading to complex orbital flip patterns and only spotty agreement with the N-body evolution.

\subsection{New systems}

\subsubsection{Circumbinary planet}
\label{sec:circumbinary}

\begin{figure}[t]
\begin{center}

\includegraphics[width=3.2in]{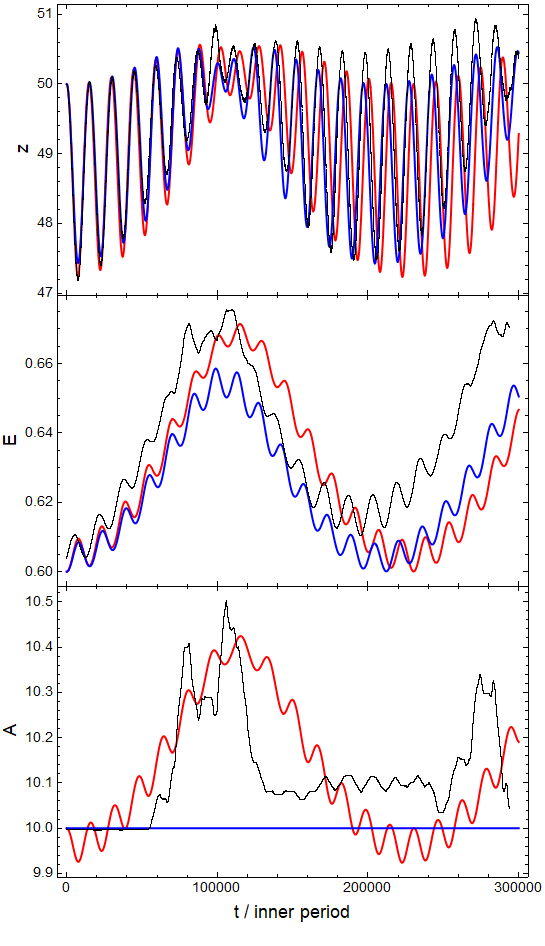}

\caption{ Evolution of the inclination, eccentricity and semimajor axis for a  circumbinary planet. The octupole, SOD and N-body solutions are represented by blue, red, and black, respectively.}

\label{fig:circumbinary} 
\end{center}
\end{figure}

In this section we focus on astrophysical systems, not explored in earlier work, that showcase the differences between octupole and SOD solutions. The first system consists of an inner binary with $m_1 = M_\odot$,  $m_2 = 10 M_\odot$, $a = 1$ a.u. and an  outer planetary body, with $m_3 =  M_\odot/1047$ and $A = 10$ a.u. The remaining initial orbital elements are
\begin{equation}
e = 0.1,\, \omega = 0^{\rm o},\, E = 0.6,\, \omega_3 = 0^{\rm o},\, z = 50^{\rm o}. \nonumber
\end{equation}
The stability factor $Y^*$ for this system is 1.368. We evolve the system for $3 \times 10^5$ periods of the inner orbit's initial Keplerian period. An astrophysical example of this system would be a planet orbiting an asymmetric stellar binary.  The inner binary in this system will be largely unperturbed by the low-mass distant body while the outer binary will be perturbed a significant amount. This type of behavior sometimes goes by the name of the inverse Kozai mechanism. The results of octupole only (blue), SOD (red), and N-body (black)  integrations are shown in Fig.\ \ref{fig:circumbinary}. The inclination and outer eccentricity plots agree reasonably well among the three simulations.  In the outer semimajor axis plot, first-order evolution through octupole order predicts a strictly constant semimajor axis, while the SOD solution and the N-body solution are in rough agreement, both predicting secular changes in $A$ with a period of about 200,000 inner orbits and an amplitude of about $0.5$ a.u.  During this evolution, the combined variations in $E$ and $A$ cause the pericenter of the planet to vary between $4$ and $3.4$ a.u.

\begin{figure}[t]
\begin{center}

\includegraphics[width=3.2in]{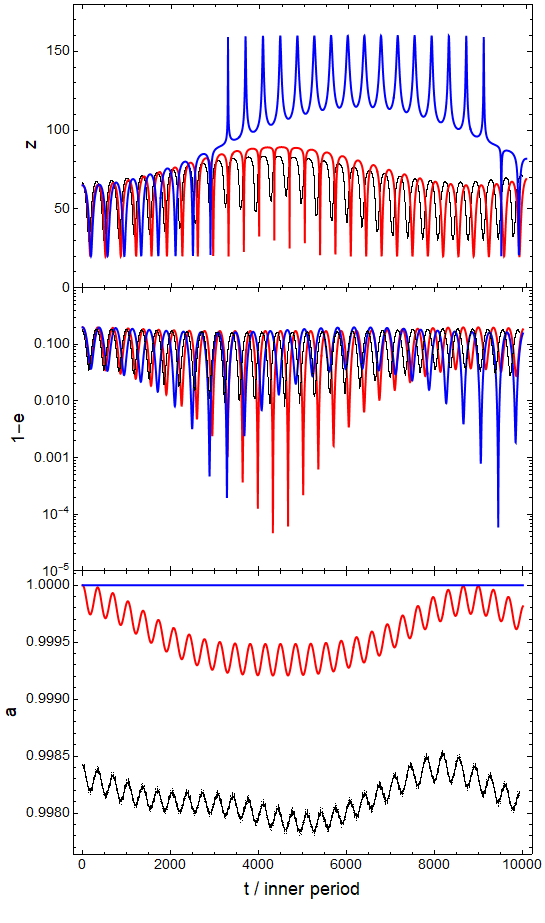}

\caption{ Evolution of the inclination, eccentricity and  semimajor axis for a  planet orbiting a member of a binary system. The octupole, SOD, and N-body solutions are represented by blue, red, and black, respectively.}

\label{fig:circumstellar} 
\end{center}
\end{figure}

\subsubsection{Planet in a binary system}

We next consider a system composed of an inner binary with $m_1 = M_\odot$,  $m_2 = M_\odot/1047$, $a = 1$ a.u. and an  outer body, with $m_3 =  M_\odot$ and $A = 14$ a.u. The remaining initial orbital elements are
\begin{equation}
e = 0.8,\, \omega = 0^{\rm o},\, E = 0.5,\, \omega_3 = 0^{\rm o},\, z = 65^{\rm o}. \nonumber
\end{equation}
The stability factor $Y^*$ for this system is 1.497. We evolve the system for $1 \times 10^4$ periods of the inner orbit's initial Keplerian period. This could represent a Jupiter-mass planet orbiting one member of a two solar mass binary. For this system, the inner binary will be perturbed much more than the outer binary, therefore we focus on the evolution of the inner elements. The results of octupole-only (blue), SOD (red), and N-body (black) integrations are shown in Fig.\ \ref{fig:circumstellar}.  The octupole solution predicts orbital flips while the SOD and N-body solutions do not. In the eccentricity plot, the N-body and SOD solutions match well, predicting eccentricities as high as $0.9999$, even without orbital flips (the averaging of the N-body data for $z$ and $e$ tends to truncate the short-timescale migrations to extreme values). In the semimajor axis plot, both SOD and N-body results agree well with respect to the amplitude and periodicity of the secular oscillations. The small vertical shift in the N-body plot most likely comes from the small difference in initial conditions arising from comparing instantaneous orbital elements to averaged orbital elements. Although the semimajor axis does vary, the amplitudes of the variations stay very small. 

\begin{figure}[t]
\begin{center}

\includegraphics[width=3.2in]{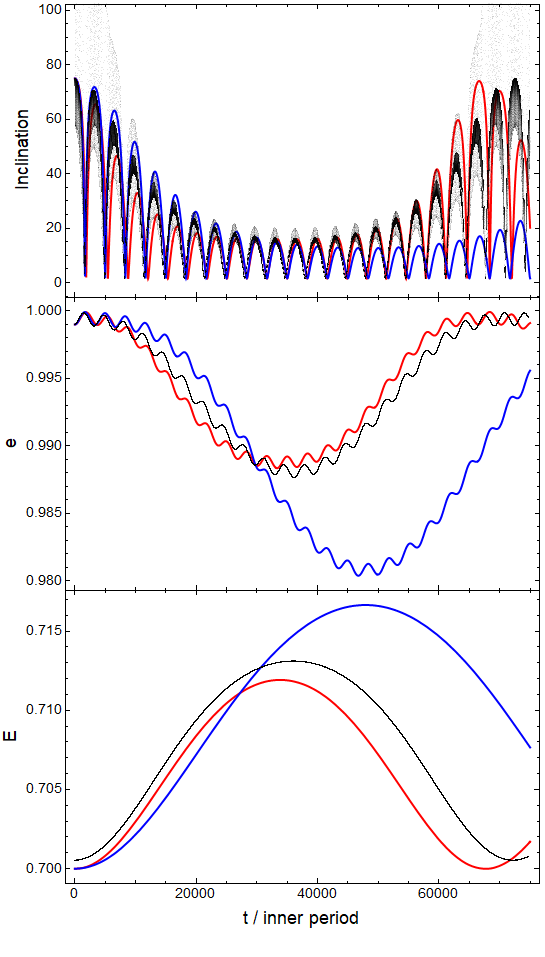}

\caption{ The evolution of the inclination, inner binary eccentricity, and outer binary eccentricity for the binary black holes example. The octupole, SOD, and N-body solutions are represented by blue, red, and black, respectively.}

\label{fig:BBHs} 
\end{center}
\end{figure}

\subsubsection{Triple black holes and merger rates}

Investigating the rate of 
binary black hole inspirals induced by a third black hole in an eccentric orbit, Su et al.\ \cite{2024ApJ...971..139S} used octupole order secular equations of motion augmented with relativistic 1PN pericenter precessions and 2.5PN gravitational radiation reaction terms to evolve a population of triple black hole systems with eccentric outer orbits. The evolution of the inner orbit's eccentricity is an important ingredient in the rate of binary mergers as the 2.5PN radiation reaction term for $da/dt$ is strongly dependent on it.  We select a system consistent with the criteria imposed in \cite{2024ApJ...971..139S} and analyze the differences between the octupole, SOD and Newtonian N-body solutions. In order to focus on these differences, we do not include relativistic effects.  The system consists of an inner binary with $m_1 = 30 M_\odot$,  $m_2 = 20 M_\odot$, $a = 100$ a.u. and a third body, with $m_3 = 30 M_\odot$ and $A = 3000$ a.u. The remaining initial orbital elements are
\begin{equation}
e = 0.999,\, \omega = 0^{\rm o},\, E = 0.7,\, \omega_3 = 0^{\rm o},\, z = 75^{\rm o}. \nonumber
\end{equation}
The stability factor $Y^*$ for this system is 1.872. We evolve the system for $7.5 \times 10^4$ periods of the inner orbit's initial Keplerian period, corresponding to about $10$ Myr. The evolutions of three different orbit elements are shown in Fig.\ \ref{fig:BBHs}. For the inner eccentricity and inclination, the three solutions are  consistent with one another for the first half of the integration, displaying conventional Kozai-Lidov oscillations. During the second half, the SOD and N-body solutions begin to diverge substantially from the octupole solution. In particular the SOD and N-body solutions show a return to large inner eccentricities much sooner than predicted by the octupole solution alone. This could lead to a significant decrease in the merger time. 
% This difference in the eccentricity solutions between octupole and dotriacontapole can impact the dynamics of the merger.
 Variations in the two semimajor  axes are tiny, of order parts in $10^3$, and it is hard to identify a definitive secular variation in the N-body data as the short timescale fluctuations dominate. 
 %This in contrast to the SOD solution that predicts variations with an amplitude 100 times larger than the N-body solution. Despite this, the periodicity of the outer semi-major axis agrees well between both solutions.
While the SOD and the averaged N-body evolutions agree in the periodicities of the variations, the N-body amplitudes are of order 100 times smaller than the SOD amplitudes.

%However, the dotriacontapole solution still predicts an amplitude and periodicity that matches the N-body solution reasonably well. For the outer semimajor axis, the variations are at the percent level. The dotriacontapole and N-body amplitudes differ by a factor of four and they appear out of phase. The periodicities match well, similar to the inner binary's semi major axis.% 

\subsubsection{The Earth-Moon-Sun system}

\begin{figure}[t]
\begin{center}
\includegraphics[width=3.2in]{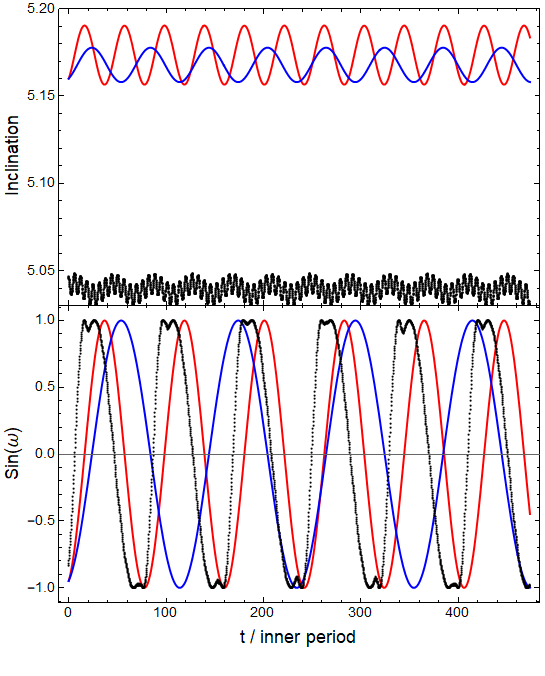}
\caption{ The evolution of the inclination and inner argument of pericenter for the Earth-Moon-Sun system. The octupole, dotriacontapole, and N-body solutions are represented by blue, red, and black, respectively.}
\label{fig:earthmoonsun} 
\end{center}
\end{figure}

The final system we consider is the triple formed from the Earth, Moon and Sun. This system is the ``ground zero'' for hierarchical triple dynamics, as Newton himself tackled it, unsuccessfully, as it turned out.  The parameters and orbital elements describing this system are 
\begin{align}
m_1 &= 3\times 10^{-6} M_\odot,\, m_2 =3.69\times 10^{-8} M_\odot,\, m_3 = M_\odot, 
\nonumber \\
a &= 0.00257 \, \text{a.u.}, \,e = 0.0554,\, \omega = 288.5^{\rm o},
\nonumber \\
A &= 1 \, \text{a.u.},\, E = 0.0167,\, \omega_3 = 85.9^{\rm o},\, z = 5.16^{\rm o}. 
\nonumber
\end{align}
The stability factor $Y^*$ for the system is 0.955. We evolve the system for $500$ periods of the inner orbit's initial Keplerian period, corresponding to $37.45$ yr. The evolution of the inclination and lunar perigee are shown in Fig.\
\ref{fig:earthmoonsun}. 
%Both dotriacontapole and octupole solutions describe the evolution of the inclination poorly, getting both the amplitude and period wrong. The inner argument of pericenter is characterized well by both dotriacontapole and octupole with the former matching the N-body better. This secular evolution of $\omega$ accounts for a fraction of the total lunar apsidal precession.
Note that the period of the octupole-order oscillations in $\sin \omega$ are off by about 30 percent from the SOD and N-body results, which agree well with each other (apart from the slight offset in the phase of $\omega$ and the value of $z$, resulting from the slight differences in initial conditions).  This can actually be seen analytically.  From Paper I, we take the first-order, quadrupole equations from Eq.\ (2.32), and the dominant, second-order, quadrupole-squared equations from Eq.\ (2.34), assume that $e \ll 1$, $E \ll 1$ and $z \ll 1$, and obtain
\begin{align}
\left (\frac{d \omega }{ d\tau} \right )_Q &\approx 3 \pi \alpha \epsilon^3 \simeq 0.053 \,,
\nonumber \\
\left ( \frac{d \omega }{ d\tau} \right )_{Q^2} &\approx\frac{27 \pi}{2} \frac{\alpha^2}{\sqrt{1+\alpha}} \epsilon^{9/2} \simeq 0.018 \,,
\label{eq:earthmoon}
\end{align}
where $\tau$ is time in units of the inner period (months).  The period associated with the first-order advance of $\omega$ is 118 months, in agreement with the blue curves in Fig.\ \ref{fig:earthmoonsun} (the oscillations in $z$ are at a frequency proportional to $2 d\omega/d\tau$).  The period of the combined first-order and second-order terms from Eqs.\ (\ref{eq:earthmoon}) is 88 months, in agreement with the N-body results, illustrating the importance of higher-order effects in the lunar motion.  The difference in the amplitudes of the oscillations of $z$ in the SOD and N-body solutions probably indicates the need to go beyond second order to get things right.  Approaches such as the Hill-Brown theory \cite{1896itlt.book.....B}  do this by expanding to high orders in perturbation theory (effectively to higher orders in $\alpha$), while exploiting the smallness of $e$, $E$, and $z$ to carry out additional expansions.

\section{Discussion}
\label{sec:discussion}

We used analytic expressions for the long-term  evolution of the averaged orbit elements of the inner and outer orbits of hierarchical triple systems derived in Paper II \cite{2024PhRvD.110h3022C} to study selected astrophysical systems.  Because three-body dynamics are fundamentally chaotic, it is difficult to make solid generalizations about the behavior even of hierarchical triples.
It is perhaps not an exaggeration in this context to paraphrase a joke from the early days of gamma-ray burst detections: If you've seen one hierarchical triple, you've seen $\dots$ one hierarchical triple.  Nevertheless, it might be possible to offer some rough generalizations.

Even in stable hierarchical triples where first-order theory and conventional wisdom assert that the averaged semimajor axes are strictly constant, second-order effects can cause substantial variations that could have interesting astrophysical consequences.

Second-order effects can play an important role in triple systems where the third body is a supermassive object.

Our SOD equations seem in many cases to give better agreement with straight N-body numerical integrations than do purely first-order equations.  This may permit investigators to perform simulations of evolutions of large numbers of systems with better fidelity than can be obtained from first-order equations alone, and with less computational burden than that needed for full N-body integrations.

\acknowledgments

This work was supported in part by the National Science Foundation
Grant PHY 22-07681.   We are grateful for the hospitality of  the Institut d'Astrophysique de Paris where part of this work was carried out.    We thank Rosemary Mardling, Hanno Rein, Alessandro Trani,  and Siyao Xu for useful discussions.  N-body simulations in this paper made use of the REBOUND N-body code \cite{rebound}. The simulations were integrated using IAS15, a 15th order Gauss-Radau integrator \cite{reboundias15}.

\bibliography{CWRefsMain,NbodyRefs}
%\bibliography{Paper3}
\end{document}